\documentstyle[preprint,aps]{revtex}
 
\begin{document} 
\title{Fermionic Symmetries: Extension of the two to one Relationship Between 
the Spectra of Even-Even and Neighbouring Odd mass Nuclei } 
\author{Larry Zamick and Y. D. Devi} 
\address{Department of Physics and Astronomy, Rutgers University, 
Piscataway\\ 
New Jersey 08854-8019, USA} 
\maketitle 
 
\begin{abstract} 
In the single j shell there is a two to one relationship between the spectra 
of certain even-even and neighbouring odd mass nuclei e.g. the 
calculated energy levels of J=0$^+$ states in $^{44}$Ti are at twice the 
energies of corresponding levels in $^{43}$Ti($^{43}$Sc) with J=j=7/2. Here an 
approximate extension of the relationship is made by adopting a truncated 
seniority scheme i.e. for $^{46}$Ti and $^{45}$Sc we get the relationship if we 
do not allow the seniority v=4 states to mix with the v=0 and v=2 states. 
Better than that, we get $\underline{very}$ close to the two to one 
relationship if seniority v=4 states are admixed perturbatively. In addition, 
it is shown that the higher isospin states do not contain seniority four 
admixtures.  
\end{abstract} 
 

\section{Introduction}

Single j-shell (j=7/2 in particular) configurations are not only simple but 
also offer ideal situations for realizing a wide variety of relationships in
the otherwise complex spectra. Although the semi-magic Ca isotopes are quite 
extensively studied \cite{CA}, the existence of such relations for open-shell 
nuclei is doubtful due to the presence of the proton-neutron interaction. 
However, it was noted by McCullen, Bayman and Zamick (MBZ) \cite{MBZ} that in 
a single j shell calculation (j=f$_{7/2}$) 
nuclei with open shells of both neutrons and protons (e.g. Scandium and 
Titanium Isotopes) there were in some cases striking relations in the 
calculated spectra of even-even nuclei and neighbouring odd-even nuclei. 
For example, the excitation energies of the J=0$^{+}$  states in $^{44}$Ti
were at twice the energies of J=j states in $^{43}$Sc (or $^{43}$Ti). It was 
further shown in MBZ technical report that the wavefunctions for the even-
even and even-odd nuclei bear a striking visual relationship. The wave 
functions for Ti were written as 

\begin{equation}
\psi^{\alpha} = \sum_{L_{p}L_{n}} D^{J_{\alpha}}\left(L_{p}L_{n}\right)
\left[\left(j^{p}\right)^{L_{p}} \left(j^{n}\right)^{L_{n}} \right]^{J}
\end{equation}

and those of Sc as

\begin{equation}
\psi^{\beta} = \sum_{L_{n}} C^{J_{\beta}}\left(L_{n}\right)
\left[j^{p} \left(j^{n}\right)^{L_{n}} \right]^{J}
\end{equation}

In the above equation $D^{J_{\alpha}}\left(L_{p}L_{n}\right)$ is the 
probability amplitude that in a state $\alpha$ with total angular momentum J 
the protons couple to L$_{p}$ and the neutrons to L$_{n}$; a similar definition 
holds for $C\left(L_{n}\right)$.

For $^{43}$Sc (J=j=7/2) and for $^{44}$Ti (J=0) the relationship is 

\begin{equation}
 D^{0}\left(L_{p}L_{n}\right) = C^{j}\left(L_{n}\right)
\end{equation}

for a state in $^{44}$Ti which is at twice the excitation energy of the 
corresponding state in $^{43}$Sc.

This relationship also holds for other pairs as well e.g. 
$\left(^{48}Ti,^{49}Ti\right)$ , $\left(^{52}Fe,^{53}Fe\right)$.

Note that the dimensions of the column vectors in the two cases is the same. 
For $^{43}$Sc J=j and the possible values of L$_{n}$ are 0,2,4 and 6. For 
$^{44}$Ti J=0 and the allowed $\left[L_{p}L_{n}\right]$ states are [0,0], [2,2],
[4,4] and [6,6]. In both cases the dimensions are the same i.e. four and four. 
This is necessary in order to have an $\underline{exact}$ two to one 
relationship.

A comparison between theory and experiment was carried out by Zamick and 
Zeng \cite{ZZ} focussing on the excitation energies of high isospin states. 
For example, for $^{48}$Ti the \underline{experimental} T=4 J=0 excitation
energy is 17.379 MeV, while in $^{49}$Ti the excitation energies of the T=7/2
J=j=7/2$^-$ state is 8.724 (the ground states have isospins T=2 and T=5/2 
respectively). The deviation from a two to one relation is -0.40\%. For 
$^{52}$Fe and $^{53}$Fe the excitation energies of the T=2 and T=3/2 states 
are respectively, 8.559 and 4.25 MeV and the percent deviation is 0.69\%. Other 
cases are considered where there is no exact two to one ratio in the theory
e.g. $^{46}$Ti and $^{47}$Ti where the percent deviation is -1.54\% and 
$^{44}$Ti, $^{46}$Ti where the percent deviation is -0.98\%. The agreement with two to one relation is surprisingly good for these pairs. However, in these 
nuclei the two to one ratio is not expected to hold for states of lower isospin.

In the following section we show the exactness of the two to one relationship 
in neighbouring pairs of nuclei with two, one particles (or holes), 
respectively, in j=7/2 shell. The approximate extension of the relation to 
certain other pairs of nuclei with more valence particles/holes is discussed 
in Sect.3 and the absence of seniority 4 contributions to the higher isospin 
states in these nuclei is shown in Sect.4. In Sect.5 higher seniority component 
admixtures are treated perturbatively. Finally, some additional remarks are 
made in Sect.6. 

\section{Exact Two to One Relation for [$^{44}$Ti, $^{43}$Sc] (and 
[$^{48}$Ti, $^{49}$Ti] and [$^{52}$Fe, $^{53}$Fe])}

The wavefunction of a Ti isotope can be written as 

\begin{equation}
\psi^{J_{\alpha}} = \sum_{L_{p}L_{n}} D^{J_{\alpha}}\left(L_{p}L_{n}\right)
\left[L_{p}L_{n} \right]^{J}
\end{equation}

where L$_p$ is the angular momentum of the two protons and L$_n$ that of the 
neutrons. We have the normalization condition

\begin{equation}
\sum_{L_{p}L_{n}} \left|D^{J_{\alpha}}\left(L_{p}L_{n}\right)
\right|^2 = 1
\end{equation}

The coefficients can be regarded as parts of a column vector representing 
the wavefunction such that 
$\left|D^{J_{\alpha}}\left(L_{p}L_{n}\right)\right|^2$ is the 
probability that in a given state $\alpha$ with angular momentum J,  the
protons couple to L$_{p}$ and the neutrons to L$_{n}$.

To obtain the wavefunction we have to diagonalize the Hamiltonian matrix, a
typical matrix element of which is 

\[\left< \left[L^{\prime}_{p} L^{\prime}_{n}\right]^{J} \left| V \right|
\left[L_{p} L_{n}\right]^{J}\right> \] 

Normally what one tries to do, and this was indeed done by MBZ \cite{MBZ}, is 
to reduce this to sums over two particle matrix elements of the form 
$\left<\left(j^2\right)^J \left|V\right| \left(j^2\right)^J \right>$
J=0,1,\ldots,7. However, we will not follow that procedure. Rather we will
first consider the matrix element of the even even nucleus $^{44}$Ti and we
will manipulate the expression so that we can get rid off the co-ordinates of 
one of the particles and thus establish a relationship  with $^{43}$Sc 
(and its mirror $^{43}$Ti). We will assume charge symmetry 
V$_{nn}$ = V$_{pp}$.

The matrix element of the even even $^{44}$Ti is written as

\[M\left(^{44}Ti\right) = \left< \left[\left(j^2\right)^{L^{\prime}_{p}}
 L^{\prime}_{n}\right]^{J} \left| V \right| 
\left[\left(j^2\right)^{L_{p}} L_{n}\right]^{J}\right> \]

We can break this into a) an interaction between the protons, b) an 
interaction between the neutrons and c) an interaction between neutrons and 
protons.

For a) and b) we get 
\[ \left[\left<L_{p}\left|V\right|L_{p}\right> + 
\left<L_{n}\left|V\right|L_{n}\right>\right]\delta_{L_{p'}L_{p}} 
\delta_{L_{n'}L_{n}} \] 

For the interaction between neutrons and protons we get

\[2 \left< \left[\left(j^2\right)^{L^{\prime}_{p}}
 L^{\prime}_{n}\right]^{J} \left| V(p;neutrons) \right| 
\left[\left(j^2\right)^{L_{p}} L_{n}\right]^{J}\right> \] 

In the above we included in V only the interaction of the second proton with 
the neutrons, and compensate by multiplying the matrix element by a factor of 
two. This is justified by the fact that the wavefunction of the two protons 
is antisymmetric.

We can use the Racah algebra to couple the second proton to the neutrons. It 
is convenient to use the unitary Racah coefficients defined by

\[ \left[[ab]j_{ab} c\right]^{J} = 
\sum_{J_{bc}} U\left(abJc;j_{ab}j_{bc}\right)
\left[a[bc]j_{bc}\right]^{J} \]

They are related to the more familiar 6j symbols by 

\[ U\left(abcd;ef\right) = (-1)^{a+b+c+d}
\sqrt{(2e+1)(2f+1)} \left\{\begin{array}{ccc}
a & b & e \\
d & c & f \end{array} \right\} \]

We get 

\begin{eqnarray}
 V_{proton-neutron} & = & \{1+(-1)^{L'-L}\} \sum_{I_{x}} 
 U\left(jjJL^{\prime}_{n}; L^{\prime}_{p}I_{X}\right)
 U\left(jjJL_{n}; L_{p}I_{X}\right) \nonumber \\
& & \left<\left[j_{p}\left[j_{p}L^{\prime}_{n}\right]^{I_{X}}\right]^J
\left| V \right| 
\left[j_{p}\left[j_{p}L_{n}\right]^{I_{X}}\right]^J\right> 
\end{eqnarray}

Since L and L$'$ are even for two protons in a single j shell the factor 
$\{1+(-1)^{L-L'}\}=2$.

We now specialize to J=0 states of $^{44}$Ti for which L$_{p}$ and L$_{n}$ are 
equal. From the unitarity condition 
\[ U\left(jj0L^{\prime}_{n}; L^{\prime}_{p} I_{X}\right)
= \delta_{I_{X}j} \]
Thus \[ V_{proton-neutron} = 2 \left<\left[jL^{\prime}_{n}\right]^{j}
\left| V \right| \left[jL_{n}\right]^{j}\right> \]
Invoking charge symmetry we find that the proton-
proton + neutron-neutron interaction equals 
2$\left<L_{n}\left|V\right|L_{n}\right>\delta_{L_{n}L_{n'}}\delta_{L_{p}L_{n}}$.
But this is just twice the corresponding matrix element between two neutrons in
$^{43}$Sc for a
state with total angular momentum J = j.  We thus see 
that a given matrix element for the J=0 state of $^{44}$Ti is twice that of the
 corresponding matrix element for the J=j state in $^{43}$Sc. Thus the column 
vectors will have identical numbers and there will be a two to one ratio for 
the energy levels. 

We show the energies and column vectors for $^{44}$Ti, $^{43}$Sc in Table.I, 
as they were originally calculated by MBZ and published in their technical 
report \cite{MBZ}. 

\vskip 0.5cm

\begin{tabular*}{6in}{c}     \hline\hline 
TABLE.I. Two to One Relation in $^{44}$Ti,$^{43}$Sc. \\ \hline\hline
Eigenvalues and Wavefunctions for J=0 levels in $^{44}$Ti \\ \hline
\begin{tabular*}{6in}{rr@{\extracolsep{\fill}}rrrr}
\multicolumn{2}{r}{Energy} & \multicolumn{1}{r}{0.0} & 
\multicolumn{1}{r}{6.5007} 
& \multicolumn{1}{r}{8.3449} & \multicolumn{1}{r}{10.8567} \\
L$_{p}$ & L$_{n}$ &  &  & T=2 & \\
0 & 0 & -0.7608 & 0.4006 & -0.5000 & 0.1037 \\
2 & 2 & -0.6090 & -0.6995 &  0.3727 & 0.0317 \\
4 & 4 & -0.2093 &  0.4156 &  0.5000 & -0.7304 \\
6 & 6 & -0.0812 &  0.4213 &  0.6009 & 0.6744 \\ 
\end{tabular*} \\ \hline
Eigenvalues and Wavefunctions for J=7/2 levels in $^{43}$Sc \\ \hline
\begin{tabular*}{6in}{rr@{\extracolsep{\fill}}rrrr}
\multicolumn{2}{r}{Energy} & \multicolumn{1}{r}{0.0} & 
\multicolumn{1}{r}{3.2503} & \multicolumn{1}{r}{4.1724} &  
\multicolumn{1}{r}{5.4284} \\
L$_{p}$ & L$_{n}$ &  &  & T=3/2 & \\
7/2 & 0 & -0.7608 & 0.4006 & -0.5000 & 0.1037 \\
7/2 & 2 & -0.6090 & -0.6995 &  0.3727 & 0.0317 \\
7/2 & 4 & -0.2093 &  0.4156 &  0.5000 & -0.7304 \\
7/2 & 6 & -0.0812 &  0.4213 &  0.6009 & 0.6744 \\ 
\end{tabular*} \\ \hline\hline
\end{tabular*}

\section{Approximate two to one Relationship for $^{46}$Ti and $^{45}$Sc}

As an example consider the pair $^{45}$Sc, $^{46}$Ti. The basis states for 
$^{45}$Sc with J=j=7/2 consist of a single proton with L$_{p}$=j and four 
neutrons with angular momenta L$_{n}$=0,2,4,6,2$^*$,4$^*$,5$^*$, where the 
states 2,4,6 have seniority $\underline{two}$ and the states 2$^*$,4$^*$ and 
5$^*$ have seniority 4. The basis states for $^{46}$Ti are [0,0], [2,2], 
[4,4], [6,6], and [2$^*$,2$^*$] and [4$^*$,4$^*$].

The dimension being 7 for $^{45}$Sc and 6 for $^{46}$Ti, it is not possible to 
have an exact two to one relationships.

Suppose, however, we make the approximation that for the lowest lying  
states we can omit the seniority $\underline{four}$ admixtures. The dimensions 
then become 4 and 4 so there is a hope for getting a two to one relationship. 
In the following paragraphs we will show that this hope is realized. 

The wavefunctions and energy levels for $^{46}$Ti and $^{45}$Sc; as calculated
by MBZ \cite{MBZ} are shown in Table.II.

\vskip 0.5cm

\begin{tabular*}{6in}{c}     \hline\hline 
TABLE.II. Approximate Two to One Relation in $^{46}$Ti,$^{45}$Sc. 
\\ \hline\hline
Eigenvalues and Wavefunctions for J=0 levels in $^{46}$Ti \\ \hline
\begin{tabular*}{6in}{ll@{\extracolsep{\fill}}rrrrrr}
\multicolumn{2}{r}{Energy} & \multicolumn{1}{r}{0.0} & 
\multicolumn{1}{r}{5.1973} 
& \multicolumn{1}{r}{7.1207} & \multicolumn{1}{r}{9.2493} 
& \multicolumn{1}{r}{11.4350} & \multicolumn{1}{r}{12.9491} \\
L$_{p}$ & L$_{n}$ &  &  & & & & T=3  \\
0 & 0 &  0.8224 & -0.3982 &  0.1527 & -0.0724 & 0.1913 & -0.3162 \\
2 & 2 &  0.5420 &  0.5245 & -0.1105 & 0.3756 & -0.3333 & 0.4082 \\
2$^{*}$ & 2$^{*}$ &  0.0563 &  0.4309 &  0.6819 & -0.5783 & -0.1082 & 0.0000 \\
4 & 4 &  0.0861 & -0.4461 & -0.2342 & -0.5244 & -0.4046 & 0.5477 \\
4$^{*}$ & 4$^{*}$ & -0.1383 & -0.4006 &  0.5755 &  0.4645 & -0.5228 & 0.0000 \\
6 & 6 & -0.0127 & -0.1454 &  0.3367 & 0.1686 & 0.6353 & 0.6583 \\ 
\end{tabular*} \\ \hline
Eigenvalues and Wavefunctions for J=7/2 levels in $^{45}$Sc \\ \hline
\begin{tabular*}{6in}{ll@{\extracolsep{\fill}}rrrrrrr}
\multicolumn{2}{r}{Energy} & \multicolumn{1}{r}{0.0} & 
\multicolumn{1}{r}{2.6204} & \multicolumn{1}{r}{3.2255} &  
\multicolumn{1}{r}{4.9559} &  \multicolumn{1}{r}{5.5225} &  
\multicolumn{1}{r}{6.4779} &  \multicolumn{1}{r}{6.6443} \\  
L$_{p}$ & L$_{n}$ &  &  &  & & & T=5/2 & \\
7/2 & 0 & 0.8210 & -0.4154 & 0.0811 & -0.0536 & 0.2068 & -0.3162 & 0.0343 \\
7/2 & 2 & 0.5434 & 0.5555 & 0.1042 & 0.1420 & -0.4362 & 0.4082 & 0.0904 \\
7/2 & 4 & 0.0846 & -0.4740 & -0.4599 & -0.0533 & -0.2079 & 0.5477 & -0.4588 \\
7/2 & 6 & -0.0130 & -0.1496 & 0.3570 & -0.0695 & 0.5429 & 0.6583 & 0.3422 \\ 
7/2 & 2$^{*}$ & 0.0428 & 0.2197 & 0.1706 & -0.9142 & 0.0160 & 0.0000 & 
-0.2912 \\
7/2 & 4$^{*}$ & -0.1462 & -0.4540 & 0.6329 & -0.0354 & -0.6030 & 0.0000 & 
0.0850 \\
7/2 & 5 & -0.0120 & -0.1319 & -0.4625 & -0.3638 & -0.2554 & 0.0000 & 0.7556 \\ 
\end{tabular*} \\ \hline\hline
\end{tabular*}

\vskip 0.5cm

For J=0 states in $^{46}$Ti the wavefunctions are of the form
\[ \psi = \sum_{L_{V}} D^{0\alpha} (L,L_{V}) [L L_{V}]^{0}\delta_{L,L_{V}} \]
i.e. the angular momentum of the two protons must equal the angular momentum 
of the four neutrons. As mentioned before there are two L=2 and L=4 states 
corresponding to seniorities v=2 and v=4.

Just as in eq.(6) in the previous section, the Hamiltonian matrix is of the 
form

\begin{eqnarray}
\left< \left[L^{\prime} L_{V}^{\prime}\right]^{0} \left| H 
\right| \left[L L_{V} \right]^{0} \right> & = & V^{L'}_{pp}\delta_{L'L} + 
V \left(f^{4}_{7/2}\right)^{L_{V}}_{\nu}\delta_{LL'} \nonumber \\
 & & + 2 U(jj0L';l'j) U(jj0l;lj)
\left<\left[j\left[jL'\right]^{j}\right]^{0} \left| V_{pn} \right|
\left[j\left[jL\right]^{j}\right]^{0}\right>
\end{eqnarray} 

The last factor is equal to $<[jL']^{j}|V_{pn}|[jL]^{j}>$ i.e. the 
proton-neutron interaction in $^{45}$Sc. The Unitary Racah coefficients are 
both equal to unity because of the zero on the left side of semicolon. 

Consider the interaction between the neutrons. At first glance it does not 
seem possible that the interaction between four neutron could equal that of 
two protons. But there is the remarkable result discussed in De Shalit and 
Talmi \cite{DS} and in Talmi's more recent work \cite{IT} that for a two-body
effective interaction in the f$_{7/2}$ shell if we limit ourselves to seniority
 v=0 and v=2 states,
\[ V\left(f^{4}_{7/2}\right)^{L}_{\nu} = 
 V\left(f^{2}_{7/2}\right)^{L}_{\nu} + constant.\]
A consequence of this result is that the seniority two states in all the 
nuclei described by $\left(f_{7/2}\right)^{n}$ configurations have nearly the 
same spectra. Therefore if we truncate to seniority 0 and seniority 2 states 
we find

\[ <[LL]^{0}|H|[L'L']^{0}> = \left(V^{L}_{pp} + V^{L}_{nn}\right)\delta_{LL'}
+ 2 <[jL']^j|H|[jL]^{j}> + constant \]

By charge symmetry $V^{L}_{nn}=V^{L}_{pp}$ and so the spectrum of $^{46}$Ti
will be double that of $^{45}$Sc provided we limit ourselves to v=0 and v=2.
This approximation should be quite good for the first few states of the two 
nuclei. We shall see in the next section that 
the situation is even better 
for states of higher isospin --- they do not have components in which the 
four neutrons couple to seniority v=4.

\section{Higher Isospin States}

In the single j shell all but one of the J=0 states in $^{46}$Ti have isospin 
T=1. The other state has isospin T=3. If we compare the wavefunction of this 
state with the higher isospin state in $^{45}$Sc we see that the numbers in 
the column vectors are the same.  Further more there are no seniority 4 neutron 
state components in the wavefunctions. 

We can explain the result as follows. $^{46}$Ti(T=3)J=0$^{+}$ is the double
analog of the J=0$^+$ groundstate of $^{46}$Ca. This nucleus has only valence 
neutrons. Thus the amplitudes D(L$_p$L$_n$)$^{J=0T=3}$ should be two particle 
fractional parentage coefficients. 

\begin{eqnarray}
^{46}Ca^{(J=0)} & = & \sum_{I_{0},v,v'} \left<\left(j^{4}\right)^{I_{0}}
v\left(j^2\right)^{I_{0}}v'\left|\left\}j^{6}J=0 v=0 \right> \right.\right.
\left[\left(j^4\right)^{I_{0}}v 
\left(j^2\right)^{I_{0}} v'\right]^{J=0 v=0} \nonumber \\
& = & \sum_{I_{0},v} \left<\left(j^{4}\right)^{I_{0}}
v\;\;\;j\left|\left\}j^{5}J=j v'=1 \right> \right.\right.
\left<\left(j^{5}\right)^{j}
v'=1\;\;\;j\left|\left\}j^{6}J=0 v=0 \right> \right.\right. \nonumber \\
& & U\left(I_{0}j(J=0)j;jI_{0}\right)
\left[\left(j^4\right)^{I_{0}}
\left(j^2\right)^{I_{0}} \right]^{J=0} 
\label{cfp}
\end{eqnarray}

Here the one particle cfp $ 
\left<\left(j^{5}\right)^{j}
v'=1j\left|\left\}j^{6}J=0 v=0 \right> \right.\right.$ is equal to one as the 
coupling of 5 particles to the sixth particle to give angular momentum zero and seniority zero state is unique and the other one particle cfp has non zero 
values only for seniority v=0 and v=2 only. Thus unlike in the previous section
 there is no necessity for truncation as only seniority v=0 and v=2 components
enter, making the relation an exact one. Once again, since J=0 the 
U-coefficient is equal to $\delta_{jI_{0}}$. Hence the 2 particle cfp
$\left<\left(j^{4}\right)^{I_{0}}v\left(j^{2}\right)^{I_{0}}v'
\left|\left\}j^{6}J=0 v=0 \right> \right.\right.$ 
is equal to the one particle 
cfp $\left<\left(j^{4}\right)^{I_{0}}v\;\;\;j\left|\left\}j^{5}J=j v'=1 \right> \right.\right.$. 
Therefore the non-zero numbers in the column 
vectors (or wavefunctions) for the T=3 and T=5/2 states are the same and they 
correspond to the non-zero values of the cfps and they can be analytically 
calculated \cite{IT,IBM} as (it should be noted that the cfps can be 
calculated to within an overall phase), 

{\large{
\[\left<\left(j^{n-1}\right)^{0}
v=0\;\;\;j\left|\left\}j^{n}J=j v'=1 \right> \right.\right. =  \sqrt{\frac 
{(2j+2-n)}{(n)(2j+1)}} \]
\[\left<\left(j^{n-1}\right)^{I_{0}}
v=2\;\;\;j\left|\left\}j^{n}J=j v'=1 \right> \right.\right. =  -\sqrt{\frac 
{2(n-1)(2I_{0}+1)}{(n)(2j+1)(2j-1)}} 
\]}}

\section{Higher Seniority Admixtures in Perturbation Theory}

The approximate two to one relationship for $^{46}$Ti and $^{45}$Sc also 
applies to the cross conjugate pair in which protons and neutron-holes are 
interchanged as well as neutrons and proton-holes. The pair in question is 
$^{50}$Cr and $^{51}$Cr. If we examine the Nuclear Data Sheets \cite{NDS}
we find that there is not sufficient data for the pair [$^{46}$Ti, $^{45}$Sc]
i.e. eventhough the T=3 0$^+$ state in $^{46}$Ti is observed at 14.153 MeV, 
the corresponding T=5/2 7/2$^-$ state in $^{45}$Sc is still missing, but there 
is for [$^{50}$Cr,$^{51}$Cr]. The T=3 - T=1 splitting in $^{50}$Cr
is 13.222 MeV and the T=5/2 - T=3/2 splitting is 6.611 MeV. This is amazing, 
the two to one relationship holds to four significant digits. 

The closeness of the results leads us to ask if we have gone as far as one can go in the previous sections. The answer is no! From Table-II we can evaluate 
the calculated percent admixtures of v=4 components in the ground states of
$^{46}$Ti and $^{45}$Sc. The respective values are 2.232\% and 2.335\%. They 
are almost the same. 

Let us therefore consider seniority 4 admixtures in perturbation theory. 
Suppose we have obtained approximate ground states for $^{46}$Ti and $^{45}$Sc
by not allowing v=4 admixtures. The approximate wavefunctions will be 

\[ \begin{array}{lccl}
^{46}Ti & \psi & = & \sum_{L;v=0,2} \tilde{D}(LL) [LL]^{0} \\
^{45}Sc & \psi & = & \sum_{L;v=0,2} \tilde{D}(LL) [jL]^{0} 
\end{array} \]

Let us consider the matrix element which couples seniority 4 admixtures in 
$^{46}$Ti

\[ M = \sum_{L'\\v=0,2} \tilde{D}(L'L') \left< \left[L'L'\right]^{0} \left|
V \right| \left[L (L v=4)\right]^{0} \right> \]

There will be no contribution from the proton-proton interaction because of 
the orthogonality of the neutron wavefunctions $<L' v \ne 4|L v = 4 > = 0$.
There will be no contribution from the neutron-neutron interaction because 
$<L' v \ne 4|V|L v = 4 > = 0$ i.e. as mentioned before seniority is a good 
quantum number for particles of one kind in the f$_{7/2}$ shell \cite{IT}.

The only contribution is from the proton-neutron interation. Using the same 
techniques as in previous sections we obtain. 

\[ M = 2\sum_{L'\\v=0,2} \tilde{D}(L'L') \left< \left[jL'\right]^{j} \left|
V \right| \left[j (L v=4)\right]^{j} \right> \]

This is exactly twice the corresponding mixing matrix element for $^{45}$Sc,
except for the fact that for $^{45}$Sc one can have L=5 v=4, but not in 
$^{46}$Ti. If we neglect the above difference we can use a 
``$V^{2}/\Delta E$'' argument. For $^{46}$Ti V and $\Delta E$ are both twice
what they are in $^{45}$Sc. The ground state energy shift $\Delta$ for 
$^{46}$Ti is

\[ \Delta \left( ^{46}Ti \right) = \frac {\left[ 
2V\left(^{45}Sc\right)\right]^{2}}{
\left[ 2\Delta E\left(^{45}Sc\right)\right]} = 2 \Delta \left(^{45}Sc\right) \]

There will be no energy shifts for the states of higher isospin because they 
have no v=4 admixtures. 

Thus in $^{45}$Sc if the unperturbed energy shift is E$_{0}$, then when v=4
admixtures are added perturbatively the shift is $E_{0} - \Delta$. In $^{46}$Ti
the shift is $2E_{0} - 2\Delta$. Thus the two to one ratio is preserved. 

In a complete matrix diagonalization there will be deviation from the 2 to 1 
ratio because of diagonal energy shifts 
and because of the previously mentioned L=5 v=4 component which is 
present in $^{45}$Sc but not in $^{46}$Ti. 

Nevertheless, the two to one ratio holds better than we would expect from 
merely truncating in seniority --- it holds when higher seniority states are 
admixed in perturbation theory. 

\section{Additional Remarks}

As mentioned in \cite{ZZ}, Rubby Sherr \cite{RS} noted that a simple 
interaction $a + b t_{1} \cdot t_{2}$ where $a$ and $b$ are constants, will 
lead to a two to one ratio for excitations of states of higher isospin, not 
only in the nuclei covered thus far but also for the pairs $^{44}$Ti, 
$^{45}$Ti and $^{46}$Ti and $^{47}$Ti. Indeed the percent deviation for these 
nuclei is small -0.98\% and -1.54\% respectively. However, for these nuclei 
there is no two to one relationship predicted for the states of lower isospin 
and the counting of states is quite different. In $^{44}$Ti there are 4 J=0 
states in the f$_{7/2}$ shell whilst in $^{45}$Ti there are 17 J=j states. The 
corresponding values for $^{46}$Ti and $^{47}$Ti are 6 and 17. 

There is one comment worth making about the seniority content of the 
J=0$^{+}_{1}$ state in $^{46}$Ti and the j=7/2$^{-}$ state in $^{45}$Sc. While
the L$_{n}$=2 v=2 probability in the states is much larger than the 
L$_{n}$=2 v=4, we find that for for L$_{n}$=4, the v=4  probability is somewhat
larger than v=2. This can be understood in terms of boson models. Roughly 
speaking, the L$_{n}$=2, v=2 state corresponds to a single d boson whereas the 
L$_{n}$=2 v=4 state corresponds to two d bosons coupled to L$_{n}$=2. It is not
surprising that one d boson admixture in the ground states should be larger 
than the two d boson admixture. 

For L$_{n}$=4 the v=2 state corresponds to one g boson whilst the v=4 state 
corresponds to two d bosons \cite{IBM}. The g boson is at about twice the
energy of the d boson and this fact causes the admixture of two d bosons to be 
comparable to the amount of one g boson in the ground state. 

\vskip 1cm 
{\bf Acknowledgements}

This work was supported by the Department of Energy Grant No.DE-FG02-95ER40940.

\end{document}